\documentclass[letterpaper]{article} 
\usepackage{aaai2027} 
\usepackage[hyphens]{url}  
\usepackage{graphicx} 
\usepackage{natbib}  
\usepackage{caption} 
\usepackage{algorithm}
\usepackage{algorithmic}
\usepackage{multirow}

\usepackage{newfloat}
\usepackage{listings}
\DeclareCaptionStyle{ruled}{labelfont=normalfont,labelsep=colon,strut=off} 
\floatstyle{ruled}
\newfloat{listing}{tb}{lst}{}
\floatname{listing}{Listing}

\usepackage{booktabs}

\title{Transpilation-Aware Runtime Prediction for Noisy Quantum Circuit Simulation}
\author{
Davud Azizov, Javier Vela-Tambo, Tian Guo
}
\affiliations{
Worcester Polytechnic Institute\\
Worcester, MA, USA\\
dazizov@wpi.edu, jvela@wpi.edu, tian@wpi.edu
}

\copyrightyear{2026}

\begin{document}

\maketitle

\begin{abstract}
Predicting the runtime of noisy quantum circuit simulations is important for scheduling, resource allocation, and performance optimization. 
However, accurate prediction is challenging because backend-aware transpilation can substantially alter the original circuit structure, while the backend-derived noise model and simulator execution behavior can introduce additional runtime variation.
We study the effectiveness of graph neural networks (GNNs) and conventional regression methods in predicting Qiskit Aer simulation runtime measured after transpilation.
We construct a dataset from a benchmark pool of 1,402 unique circuits spanning 22 circuit families, two Qiskit fake-backend configurations, and four transpiler optimization levels.
Specifically, we compare a source GNN using original circuit information, hybrid GNN combining source-level graph with post-transpilation features, and transpiled GNN using only transpiled circuit information, along with five regression models. In the overall-model setting, the transpiled GNN achieves the strongest performance among the graph-based representations at all four optimization levels, obtaining $R^2$ values of 0.974, 0.713, 0.745, and 0.605 for optimization levels 0 through 3, respectively. However, under backend-specific evaluation, the advantage of GNN decreases, with conventional regression models matching or outperforming the GNNs in several settings. 
These results indicate that post-transpilation information is useful, while the value of explicit graph modeling depends on the backend and optimization level.

\end{abstract}

\section{Introduction}

Quantum computing has emerged as a promising paradigm for solving computationally challenging problems, with recent advances indicating the potential for quantum advantage in specific practical tasks~\cite{quantumadvantage}. With the growing availability of cloud-accessible quantum hardware~\cite{ravi_quantum_2021}, noisy simulators play a complementary role in developing and evaluating quantum workloads before hardware execution. By modeling backend-specific constraints and noise in a controlled environment, simulators support circuit debugging, compiler evaluation, benchmarking, and repeated experimentation without requiring continual access to a quantum device. However, these simulations can themselves be computationally expensive, and their runtimes can vary substantially across circuits and backend configurations. Runtime and resource estimates support job scheduling, resource allocation, and performance optimization in quantum operating systems and schedulers~\cite{QOS,Giortamis2025Other,Ravi2021Other-01}; for simulation workloads, they can help set per-job time limits and estimate the computing resources required for large batches of circuit simulations. Motivated by these needs, we study the prediction of noisy-simulator runtime across quantum circuits and backend configurations.

Predicting noisy-simulator runtime is challenging because the simulator
does not execute the original high-level circuit directly. We refer to
this original circuit before transpilation as the \emph{source-level
circuit}. Before simulation, the source-level circuit is transpiled for
a target backend configuration, which requires its operations to be
decomposed and routed according to the backend's native gate set and
coupling map~\cite{transformation}. The resulting \emph{transpiled
circuit} is the backend-compatible circuit processed by the simulator.
Transpilation can substantially change circuit depth, gate composition,
and the number of two-qubit operations. The backend's noise
configuration can introduce additional runtime variability during
simulation. Consequently, the relationship between source-level circuit
properties and simulator runtime is indirect, nonlinear, and
backend-dependent.

Prior work has examined quantum circuit performance through
benchmarking~\cite{quetschlich2023mqtbench,supermarq},
resource estimation~\cite{suchara_qure_2013,dam_using_2024}, and
backend-aware compilation~\cite{murali_noise-adaptive_2019,
quetschlich_mqt_2025}. Benchmarking and resource-estimation studies
provide tools for characterizing circuits and estimating hardware
requirements, while backend-aware compilation focuses on mapping
circuits to specific devices. Most closely related, \citet{ma_understanding_2025} propose a
graph-transformer model~\cite{velickovic_graph_2018} for execution-time
prediction on simulators and real hardware. Their approach primarily
uses source-level, pre-transpilation circuit representations and does not explicitly examine either the predictive value of backend-specific features extracted after transpilation or the transpiled circuit graph itself. Therefore, it remains unclear how much transpilation-aware information improves noisy-simulator runtime prediction across different transpiler optimization levels and backend settings.

To address this gap, we present \textbf{Q-TARP} (\textbf{Q}uantum \textbf{T}ranspilation-\textbf{A}ware \textbf{R}untime \textbf{P}rediction),
a framework for predicting noisy quantum simulator runtime using transpilation-aware circuit representations.
We build on the two-branch graph and global-feature architecture of \citet{ma_understanding_2025} and propose three representation settings.

Given a quantum circuit and a target backend, our goal is to predict the simulation execution time after backend-specific transpilation at a fixed shot count. We use the model of \citet{ma_understanding_2025}, which uses the original pre-transpilation circuit graph and source-level global features, as our baseline and refer to it as the \emph{source GNN}. We compare this baseline against two proposed transpilation-aware variants: a \emph{hybrid GNN}, which retains the source-level circuit graph but augments the source-level feature vector with 13 post-transpilation global features, and a \emph{transpiled GNN}, which uses the backend-specific transpiled circuit graph together with transpilation-aware global features.

To understand the value of explicit graph topology modeling, 
we additionally compare these models with five machine learning (ML) regression methods: linear regression, ridge regression, SVR, random forest, and XGBoost. To evaluate these predictors, we construct a circuit dataset that consists of source-level and transpiled graphs for four transpilation optimization levels and two simulation backends using Qiskit Aer simulator~\cite{javadi-abhari_quantum_2024}. In the overall-model setting, the transpiled GNN provides the strongest and most consistent results across all optimization levels, outperforming both source and hybrid GNNs, as well as ML models.  
In summary, our key contributions are:
\begin{itemize}

\item \textbf{Circuit representation and predictive modeling.}
We formulate and evaluate three GNN graph representations (source, hybrid, and transpiled), alongside five ML regression methods, and quantify the effects of feature representations and model choices on noisy-simulator runtime prediction.

\item \textbf{Backend and optimization-level analysis.}
We evaluate the models across two backend configurations and four optimization levels, showing when transpiled circuit information and optimization levels provide good trade-offs in prediction accuracy and overhead.

\item \textbf{Open noisy-simulator dataset and reproducible pipeline.}
We release a noisy-simulator runtime dataset and evaluation pipeline for all representations, spanning two
backend configurations and four transpiler optimization levels. \footnote{\textcolor{magenta}{\url{https://github.com/cake-lab/q-tarp}}}
\end{itemize}

\section{Background}
\paragraph{Quantum Circuit.} A quantum circuit represents a quantum algorithm as a sequence of operations applied to quantum bits (qubits). These operations include quantum gates that transform the state, and measurement operations that extract classical results.

\begin{figure}[t]
  \centering
  \includegraphics[width=\columnwidth]{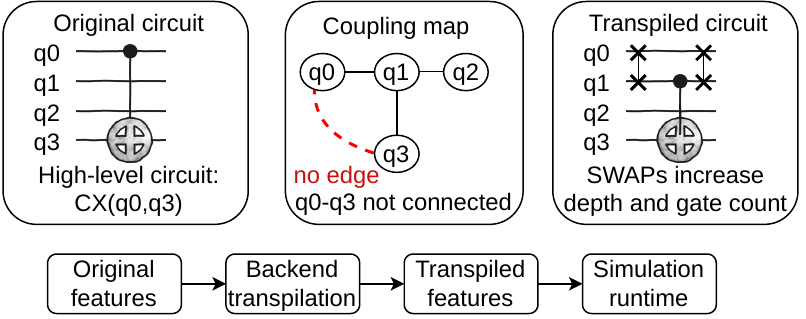}
  \caption{Transpilation transforms a source-level circuit into a backend-compatible form by decomposing gates and inserting SWAP operations. 
  Circuits were created with the Quantum Circuit Library~\cite{wilkens_quantum_circuit_library}.}
  \label{fig:transpilation}
\end{figure}

\paragraph{Circuit Compilation.} Before execution, a quantum circuit is transpiled to satisfy the constraints of the target backend. Transpilation rewrites the circuit using the backend's native gate set, maps logical qubits to physical qubits, and adapts two-qubit operations to the backend coupling map. Because hardware devices support only specific native gate sets, any gates that are not supported by the backend are decomposed into equivalent sequences of native operations~\cite{quetschlich2023mqtbench}. Additionally, two-qubit gates between logical qubits that are physically unconnected, as defined by the coupling map of the device, are resolved by inserting SWAP operations, a step known as routing. This process produces a backend-compatible circuit that is functionally equivalent to the source-level circuit but structurally different. In Qiskit~\cite{javadi-abhari_quantum_2024}, transpilation can be performed with different optimization levels, which control how much effort the transpiler spends on rewriting and simplifying the circuit. Higher optimization levels may reduce circuit depth or gate count, but they can also increase compilation time.
Compilation latency can itself be a substantial component of end-to-end
processing time, motivating both high-performance transpilation frameworks
and techniques that amortize or avoid repeated compilation
\cite{Hua2023Other-01,Quetschlich2023Other-01}.
Figure~\ref{fig:transpilation} illustrates the transpilation process at a high level.

\begin{figure*}[t]
  \centering
\includegraphics[width=\textwidth]{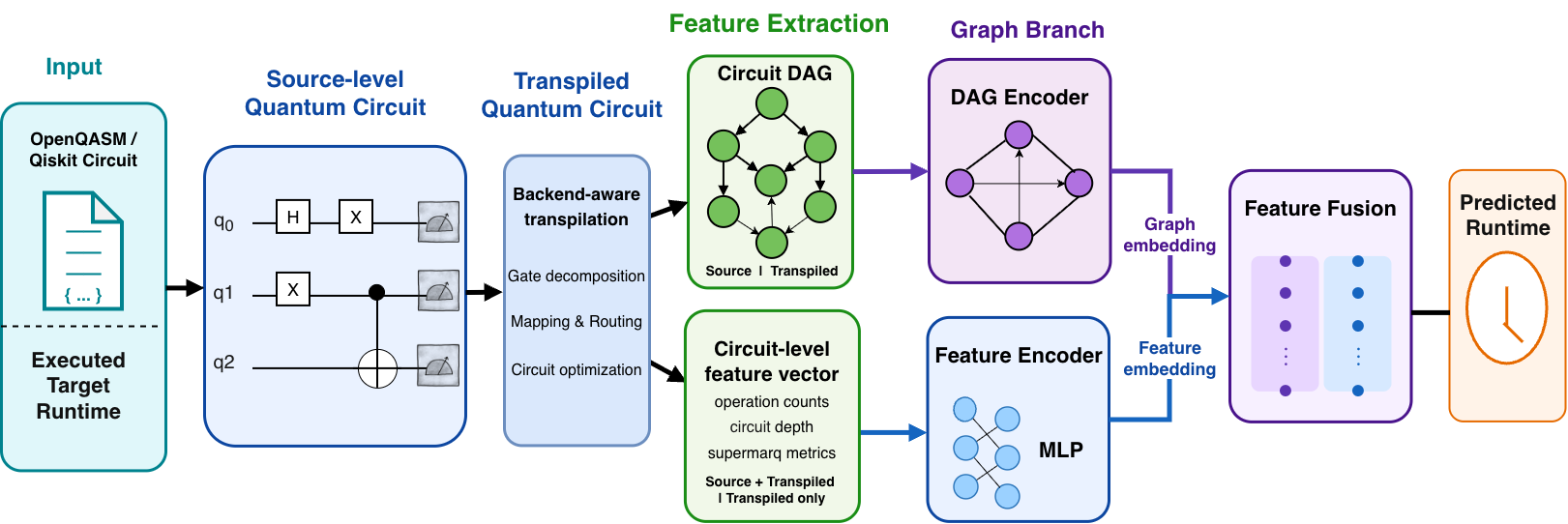}
  \caption{Our high-level end-to-end GNN prediction pipeline.
  }
  \label{fig:architecture}
\end{figure*}

\section{Related Work}

\paragraph{Quantum Benchmark.}

MQT Bench~\cite{quetschlich2023mqtbench} provides a large benchmark suite
with circuits represented at multiple abstraction levels, supporting reproducible evaluation across the quantum software stack. SupermarQ~\cite{supermarq} extends benchmark design with circuit-level characterization metrics such as program communication, critical depth, entanglement ratio, parallelism, and liveness. These metrics are useful because circuit execution cost is not determined only by the number of qubits or gates, but also by circuit structure, connectivity, depth, and parallelism.
Recent work further shows that circuit depth alone is not an accurate proxy for runtime prediction because operations may have different execution costs~\cite{Tremba2025Other}, motivating richer gate- and backend-aware representation and metrics.
Our work builds on this insight by using such circuit-level features, extracted from both the source and transpiled circuit, as inputs to a machine learning model for predicting noisy-simulator execution time.

\paragraph{Resource Estimation.}
The primary goal of resource estimation frameworks is to calculate the physical cost, including qubit count and execution time required to run the algorithms on fault-tolerant quantum computers. Frameworks like QuRE~\cite{suchara_qure_2013} and Azure Quantum Resource Estimator~\cite{dam_using_2024} provide estimates that help compare how different hardware assumptions, error-correction schemes, and algorithm designs affect the resources needed for execution. 

These tools estimate the resources required to execute algorithms on fault-tolerant quantum hardware,
whereas our work predicts measured runtime on noisy quantum simulators.

\paragraph{Backend-Aware Compilation and Selection.}

\citet{murali_noise-adaptive_2019} propose noise-adaptive, calibration-aware compiler techniques for Noisy Intermediate-Scale Quantum (NISQ) devices. MQT Predictor~\cite{quetschlich_mqt_2025} extends this direction by selecting a suitable quantum device and performs device-specific compilation for a given quantum algorithm. 
Compiler configuration selection has also been formulated as a supervised prediction problem to select promising combinations of devices, compilers, and compiler settings for a given circuit~\cite{Quetschlich2023Other}. More recently, \citet{villar_ai_2025} applied reinforcement learning to permutation-circuit synthesis across different hardware topologies, demonstrating generalization to previously unseen topologies.
These approaches optimize backend selection and compiled-circuit quality, whereas our work predicts 
the runtime of transpiled circuits on a noisy simulator.

\paragraph{Quantum Circuit Runtime Analysis.}
QSimBench~\cite{bisicchia_qsimbench_2025} provides an execution-level benchmark suite for noisy quantum simulation. It provides a comprehensive suite of over 20 million shot-level execution traces across different algorithms and multiple circuit sizes, using both ideal and noisy simulation backends. This work provides a standardized foundation for evaluating Quantum Software Engineering (QSE) tools. While QSimBench simplifies research by removing the need for repeated, computationally expensive simulations, our work complements this by predicting the execution time of the circuits. 
The closest work, by \citet{ma_understanding_2025}, combines graph-based circuit structure and global features in a graph-transformer model evaluated on noisy simulators and real hardware. Building on this design, we incorporate new source-level features and backend-specific post-transpilation features and analyze their effects across optimization levels and backends using a reproducible dataset and preprocessing pipeline.
\paragraph{Quantum Circuit Simulation.}
Classical simulation plays an important part in quantum computing. 
\citet{Cicero} emphasize that, given the limited resources and availability of current quantum hardware, classical simulators remain one of the most practical ways to develop and test quantum algorithms. Simulation cost, however, depends strongly on backend, simulation method and circuit characteristics. \citet{VALLERO2026107927} show that circuit structure can considerably affect the relative performance of some simulators, and circuit-level analysis can help choose the right simulation.
\citet{Guerreschi2022fastsimulationof} shows that simulator-specific circuit optimization can substantially reduce quantum circuit simulation time. Motivated by these observations, our work focuses on modeling noisy quantum simulation runtime and investigates whether circuit characteristics introduced through backend-aware transpilation provide additional information for execution-time prediction.

\section{Q-TARP Design}

The goal of this study is to predict the Qiskit Aer simulation runtime~\cite{javadi-abhari_quantum_2024} of a quantum circuit after it has been transpiled for a target backend, using a fixed shot count.
We formulate this as a supervised regression problem, where each training example consists of a circuit and its measured simulator runtime. As shown in Fig.~\ref{fig:architecture}, each circuit is encoded using two complementary representations, which are processed by a two-branch neural network to estimate the runtime.

\subsection{Circuit Representation and Feature Extraction}
\label{subsec:circuit_representation}

We use three complementary representations for each circuit. In the \emph{source setting}, which follows the representation used by \citet{ma_understanding_2025}, the graph branch uses a source-level directed acyclic graph (DAG) extracted from the original circuit before transpilation. In the DAG, nodes represent operations and edges capture dependencies, allowing the model to learn circuit structure before backend-specific transpilation effects are introduced. The accompanying global feature vector for the MLP branch contains 41 source-level features.  The source-level features include 36 features following \citet{ma_understanding_2025}, together with five SupermarQ metrics~\cite{supermarq} that capture higher-level structural properties, such as communication, parallelism, and liveness, beyond basic gate counts and circuit depth.

In the \emph{hybrid setting}, we keep the same source-level DAG but extend the global feature vector with 13 additional transpilation-aware features. The transpilation-aware features are computed after mapping the circuit to a target backend. They include the transpiled circuit depth, total operation count, two-qubit operation count, backend-native gate counts, and the five SupermarQ metrics recomputed on the transpiled circuit. These features capture backend-specific changes introduced by transpilation.

In the \emph{transpiled setting}, the graph branch is built directly from the backend-specific circuit rather than the original circuit. The global feature vector is also computed from the transpiled circuit itself, using the same 41-feature setting. In this design, both branches are built entirely from the transpiled circuit.

\subsection{Two-Branch Graph Regression Model}
\label{sec:prediction_model}

The GNN-based predictor follows a two-branch graph regression design similar to
that of \citet{ma_understanding_2025}. The graph branch encodes the circuit DAG, while the MLP branch encodes the global feature vector. Their embeddings are concatenated and passed through fully connected layers to predict the noisy-simulator runtime.
In our implementation, the graph branch applies three \texttt{TransformerConv} layers to the DAG node features, followed by global mean pooling to obtain a graph-level embedding. The global-feature branch uses a two-layer multilayer perceptron. 
We refer to the model trained in the source, hybrid, and transpiled settings as \emph{source GNN}, \emph{hybrid GNN}, and \emph{transpiled GNN}, respectively.

\section{Evaluation}

\subsection{Evaluation Methodology}

Our evaluation of Q-TARP examines how transpilation-aware features, transpiled graphs, and backend-specific predictors affect simulator-runtime prediction accuracy. 
We ran Qiskit 0.44.2 with Qiskit Aer 0.11.2 simulator for different optimization levels and backends on HPC compute nodes with 32 allocated CPU cores and 256 GB RAM, to obtain the transpiled circuit graphs and the simulator execution times.

\begin{figure}[t]
  \centering
\includegraphics[width=\columnwidth]{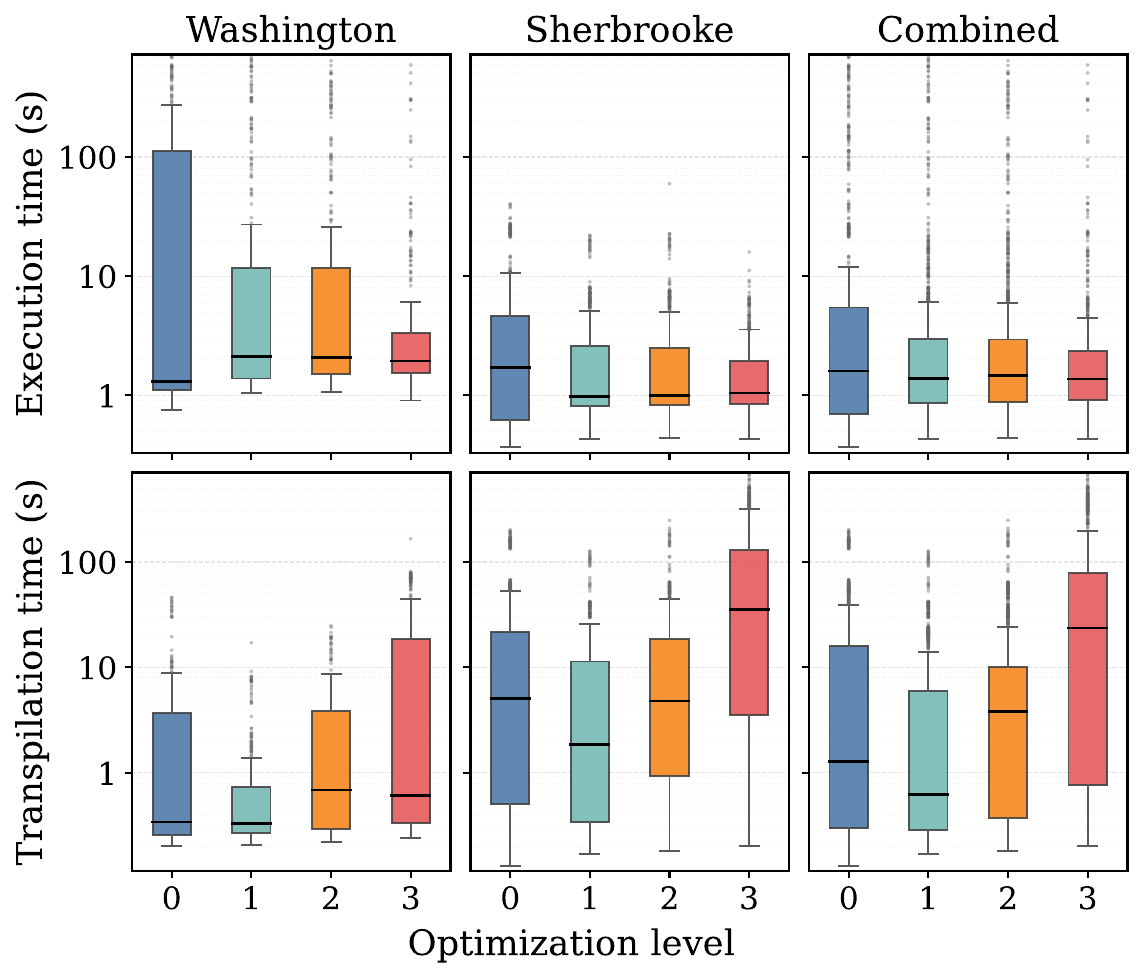}
  \caption{Execution and transpilation time distributions for each backend and for the combined dataset across all optimization levels. The top row shows execution-time distributions, while bottom row shows transpilation time.
  }
  \label{fig:distribution}
\end{figure}

\subsubsection{Dataset Construction.}
We start from 1510 original quantum circuits for each backend from \citet{ma_understanding_2025},
which are derived from MQT Bench~\cite{quetschlich2023mqtbench}, spanning 22 circuit families. After removing duplicate circuits present in the original dataset, 1402 unique circuits remain for each backend. Each
circuit is paired with both \texttt{FakeWashington} and
\texttt{FakeSherbrooke}, transpiled separately at four Qiskit optimization levels (Opt 0, Opt 1, Opt 2, Opt 3), and executed for 1024 shots using Qiskit Aer with device noise based on the corresponding fake backend's calibration data. Each circuit-backend execution that completes within the 900-second time limit forms one sample, with the reported simulator runtime as the prediction target.
Table~\ref{tab:datasetSize} shows how resulting samples are distributed across backends and optimization levels.
Transpiled setting contains fewer samples because we cap the circuit DAG to be no more than 300K nodes due to the long DAG construction time. To control for this difference, we additionally perform a controlled comparison in which all three GNN representations use the same set of circuits and identical train-test splits.

\begin{table}[t]
\centering

{\small
\setlength{\tabcolsep}{3pt}

\begin{tabular}{llcccc}
\toprule
\textbf{Representation} & \textbf{Backend}
& \textbf{Opt 0} & \textbf{Opt 1}
& \textbf{Opt 2} & \textbf{Opt 3} \\
\midrule

Source / Hybrid
& Washington & 270 & 359 & 367 & 297 \\
& Sherbrooke & 838 & 783 & 788 & 762 \\
& \textbf{Total}
& \textbf{1108} & \textbf{1142}
& \textbf{1155} & \textbf{1059} \\

\midrule

Transpiled
& Washington & 270 & 359 & 367 & 297 \\
& Sherbrooke & 473 & 687 & 697 & 705 \\
& \textbf{Total}
& \textbf{743} & \textbf{1046}
& \textbf{1064} & \textbf{1002} \\

\bottomrule
\end{tabular}
}
\caption{Dataset size by backend, optimization level, and circuit representation.}
\label{tab:datasetSize}

\end{table}

Fig.~\ref{fig:distribution} summarizes resulting execution and transpilation time distributions across optimization levels and backends. The Washington backend has the largest execution time distribution at Opt 0, with higher optimization levels mainly reducing the upper tail of the distribution. Sherbrooke shows a more compact execution time distribution, although reductions at higher optimization levels are still clear. The third column shows the combined distribution of both backends. 
Transpilation, on the other hand, follows the opposite trend. Opt 3 generally has the largest median and upper-range transpilation times, particularly on Sherbrooke. Overall, the figure~\ref{fig:distribution} highlights a clear trade-off: more aggressive optimization may reduce the execution times, but it introduces larger transpilation overhead. 

\subsubsection{Data Splitting and Model Selection.}
We generate 1,000 candidate 80/20 train-test splits and select the one
that minimizes a distribution-matching score based on backend, circuit
family, and execution-time distributions. The split is selected before model training without using model predictions or test-set performance, and the resulting test partition is held out until final evaluation, ensuring that all source circuits in the test set are completely unseen during training.  Global-feature dimensions that are zero across
the training samples are removed, and the remaining dimensions are
standardized feature-wise using statistics computed only from the
training split. Hyperparameters are selected using five-fold
cross-validation on the training split. Using the configuration with
the best mean validation performance, we retrain the model on the full
training split and evaluate it on the held-out test set. For the transpiled setting, experiments use the same corresponding dataset or a subset of it after circuits are removed by DAG-size filtering. 

\subsubsection{ML baseline models.}
We compare the proposed graph-based predictors against five ML regression baselines using the same global feature vector. These models cover linear methods (linear and ridge regression), kernel-based regression (support vector regression (SVR) with an RBF kernel~\cite{Smola2004}), and tree-based methods (random forest and XGBoost~\cite{10.1145/2939672.2939785}). Hyperparameters are selected through cross-validation over predefined search spaces. 

\subsubsection{Evaluation Metrics.}
We evaluate prediction performance using two standard regression metrics: root mean squared error (RMSE) and the coefficient of determination ($R^2$). RMSE measures prediction error in the same units as simulator runtime, with lower values indicating greater accuracy. $R^2$ measures the proportion of runtime variance explained by the model, with higher values indicating a better fit.

\subsection{Experimental Results}

We evaluate how the proposed Q-TARP hybrid and transpiled GNN-based predictors perform compared to the source GNN, based on \citet{ma_understanding_2025}, and commonly used ML models. 
First, we study the overall model prediction setting: a model trained on circuits executed on both noisy simulator backends. This setting allows us to evaluate whether transpilation-aware features and transpiled graphs improve prediction over source-level features, and whether the improvement depends on the transpilation optimization level. Second, we compare backend-specific prediction with backend-stratified results from the overall model. This allows us to test whether separate training for each backend is necessary and whether the benefit of transpilation is consistent across backends.  
Additionally, we evaluate trade-offs of optimization level on transpilation overhead, and GNN inference accuracy and time. 

\subsubsection{Prediction Comparison Among Models.}

\begin{figure}[t]
  \centering
\includegraphics[width=\columnwidth]{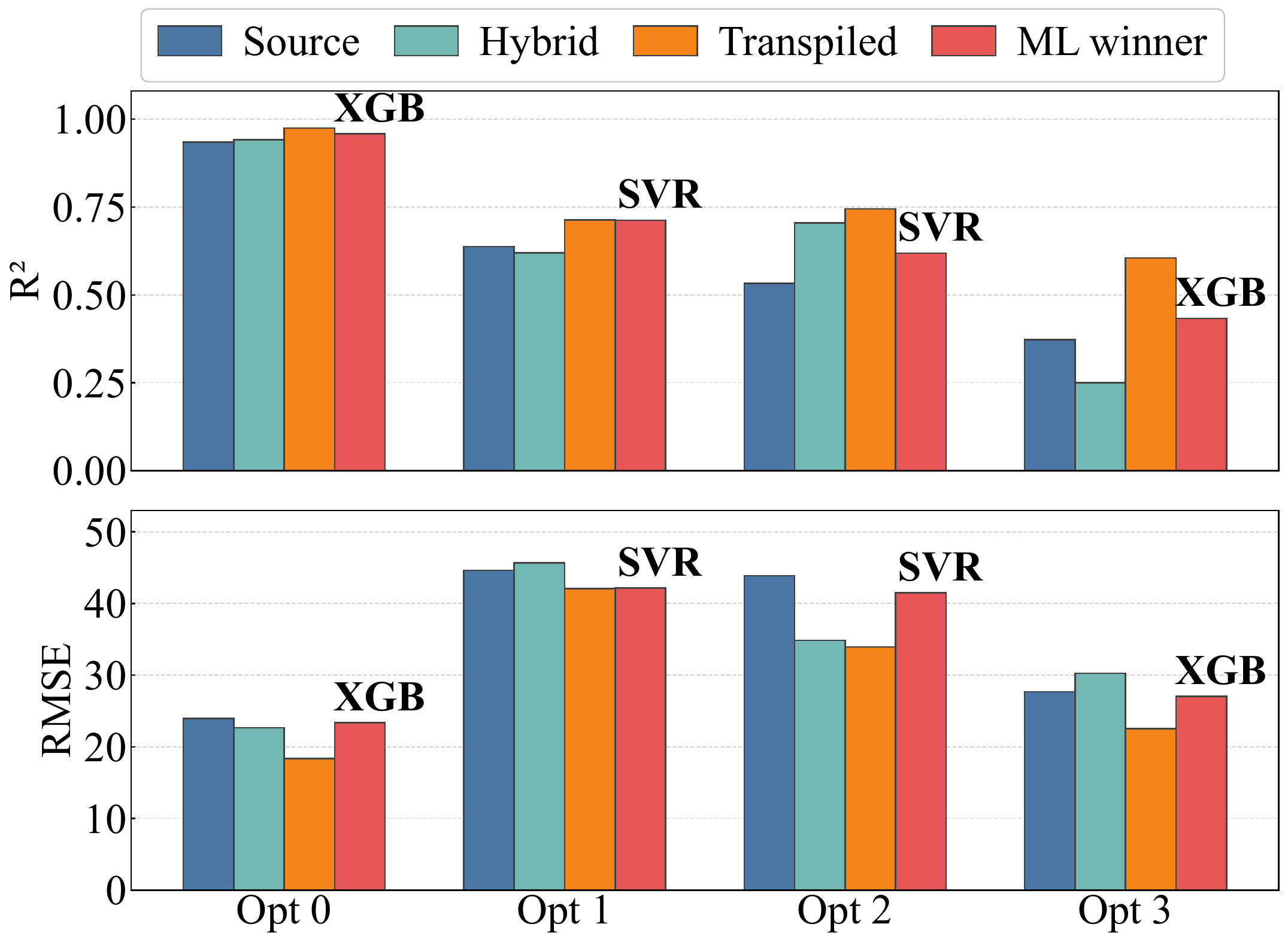}
  \caption{Overall performance of GNN models and the best ML model across optimization levels.
  }
  \label{fig:gnns}
\end{figure}

\begin{table}[t]
\centering
{\small
\setlength{\tabcolsep}{2pt}

\begin{tabular}{lcccccccc}
\toprule
& \multicolumn{2}{c}{\textbf{Opt 0}}
& \multicolumn{2}{c}{\textbf{Opt 1}}
& \multicolumn{2}{c}{\textbf{Opt 2}}
& \multicolumn{2}{c}{\textbf{Opt 3}} \\
\cmidrule(lr){2-3}
\cmidrule(lr){4-5}
\cmidrule(lr){6-7}
\cmidrule(lr){8-9}

\textbf{Model}
& $R^2$ & RMSE
& $R^2$ & RMSE
& $R^2$ & RMSE
& $R^2$ & RMSE \\
\midrule

Linear
& .786 & 52.85
& .590 & 50.33
& .578 & 43.64
& .082 & 34.38 \\

Ridge
& .787 & 52.72
& .586 & 50.56
& .582 & 43.44
& .093 & 34.16 \\

SVR
& .690 & 63.59
& \textbf{.712} & \textbf{42.16}
& \textbf{.619} & \textbf{41.50}
& .307 & 29.87 \\

RF
& .944 & 26.97
& .640 & 47.13
& .613 & 41.83
& .402 & 27.74 \\

XGB
& \textbf{.958} & \textbf{23.39}
& .605 & 49.40
& .430 & 50.72
& \textbf{.433} & \textbf{27.02} \\
\bottomrule
\end{tabular}
}

\caption{Overall-model performance of five regression models using
transpilation-aware global circuit features across all optimization levels.}
\label{tab:baseline_results}
\end{table}

Fig.~\ref{fig:gnns} evaluates the three GNN representations across four optimization levels using $R^2$ and RMSE on the overall model prediction setting. Since the Transpiled GNN is the strongest of the three GNN variants, we compare it with conventional ML baselines using the corresponding transpiled  representation. Table~\ref{tab:baseline_results} reports the results for linear regression, ridge regression, random forest, SVR, and XGBoost under this setting, and the best-performing baseline at each optimization level is included in Fig.~\ref{fig:gnns}. 

The Transpiled GNN provides the strongest and most consistent results across all optimization levels, outperforming both the Source and Hybrid GNNs as well as the baseline winner. Its gain is relatively small  at Opt 0 and 1, where transpilation applies relatively limited optimization. This shows that the original representation can still retain much of the information needed for the prediction. The benefit of the Transpiled GNN becomes clearer at higher optimization levels, Opt 2 and 3, where transpilation phase has more optimizations, such as more intensive layout, routing, cancellation and resynthesis passes~\cite{qiskit_transpiler}. These results demonstrate that simply adding transpilation-aware global features to the source graph is not enough to capture all of the changes introduced during compilation. Directly modeling the post-transpilation graph provides a representation that is closer to the circuit actually executed by the simulator and becomes especially useful under stronger optimization.

Since the transpiled representation excludes circuits whose DAGs exceed the 300K-node threshold, its dataset is smaller than those of the Source and Hybrid GNNs. To verify that the observed performance differences are not caused by this difference in dataset composition, we additionally perform a controlled comparison  using the same set of circuits and identical train-test splits for all three GNN representations. The relative performance remains consistent with the main results, with the Transpiled GNN achieving the strongest predictive performance in all above cases. This provides additional evidence that its advantage is due to the post-transpilation representation rather than differences in dataset.

A second pattern is that $R^2$ generally decreases as the optimization level increases. This does not mean that transpiled representation becomes less useful. Instead, as shown in Fig.~\ref{fig:distribution}, the Opt 3 dataset has a much tighter runtime distribution, with many circuits having similar execution times. Because $R^2$ measures prediction error relative to the variance of actual runtimes, smaller variance makes it harder to achieve a high $R^2$ even with low absolute errors. Consistent with this,  RMSE at Opt 3 is lower, despite the reduced $R^2$.

\subsubsection{Backend-Specific Predictions}

\begin{figure*}[t]
\centering
\includegraphics[width=\textwidth]{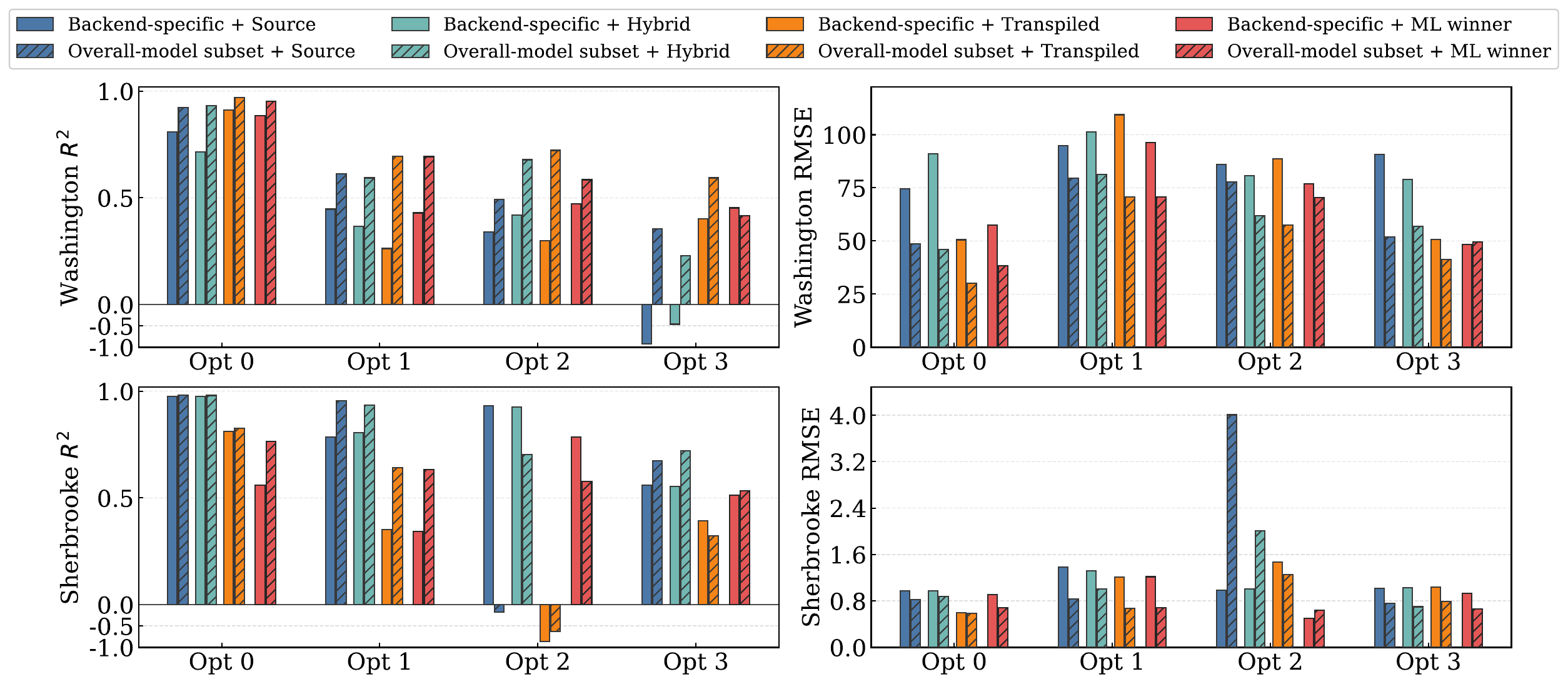}
\caption{Backend-specific and overall-model performance for GNN models across optimization levels.
}
\label{fig:backend_results}
\end{figure*}

Fig.~\ref{fig:backend_results} compares the separately trained backend-specific models (solid bars) with the backend-stratified results of the overall model (hatched bars). Because Washington and Sherbrooke have substantially different runtime ranges, 0.75--689.71\,s and 0.37--59.69\,s, respectively, absolute RMSE values should be interpreted within rather than across backends.

\textbf{Washington.} 
Washington shows a clear benefit from backend-stratified results. Across optimization levels and all three representations, the overall Transpiled GNN is the best-performing model for both $R^2$ and RMSE, outperforming the backend-specific GNNs and the best ML models. However, ML models remain competitive. XGBoost is the best ML model at Opt 0 and Opt 3, while SVR is best at Opt 1 and
Opt 2, with none performing better than the overall Transpiled GNN.

One possible reason behind this behavior is Washington's much broader runtime distribution. Washington contains fewer successful executions than Sherbrooke in overall dataset because of its outliers, as shown in Fig.~\ref{fig:distribution}. This is also visible in the prediction errors. For the overall Transpiled GNN, the 90th percentile absolute error is $51.8$, $128.1$, $58.6$, and $5.9$ at optimization levels from 0 to 3, respectively. Notably, Opt 3 reduces this error to $5.9$, indicating the importance of optimization in prediction. 
Overall, these results suggest that separate training is not necessary for Washington.

\textbf{Sherbrooke.} A different pattern exists for Sherbrooke in terms of $R^2$. Unlike Washington, the overall model does not consistently outperform backend-specific training and none of the representations is consistently better across optimization levels. 

However, RMSE gives a pattern similar to Washington. Within the overall model setting, the Transpiled GNN generally achieves the lowest RMSE among GNN representations. At Opt 2, however, the backend-specific XGBoost is the clear winner with an RMSE of $0.50$~s, while at Opt 3 the overall-model Random Forest achieves the lowest RMSE at $0.66$~s. 

One possible explanation for this difference is that Sherbrooke has much tighter runtime distribution than Washington, so relatively small absolute errors can still noticeably reduce $R^2$. Consistent with this, the 90th percentile absolute error remains below approximately $1.25$~s in all settings. Overall, Sherbrooke shows that backend-specific training can improve prediction in selected settings, including for the Transpiled GNN at some optimization levels, although no representation consistently benefits from separate training across all settings.

\subsubsection{Optimization Level Impact on Transpilation.}

To examine when additional optimization becomes worthwhile, we define \emph{total runtime} as the sum of transpilation time and execution time, and analyze only the circuits successfully completed at all optimization levels. We analyze two workload characteristics independently: circuit depth and Opt 0 execution time. For each characteristic, circuits are ranked separately and divided into four approximately equal-sized groups. Thus, for depth, group 1 contains the shallowest circuits, while group 4 contains the deepest; while for Opt 0 execution time, group 1 contains the fastest executions, and group 4 the slowest. Within each group, we compare optimization levels from 1 to 3 against 0 for the same circuits and report the percentage for which the higher optimization level produces a lower total runtime. 

Table~\ref{tab:tradeoff} reports the results for Opt 0 execution time and circuit depth. It shows that the benefit of additional optimization depends on the circuit characteristics. For example, on Washington, the percentage of circuits benefiting from Opt 1 increases from $15.2\%$ to $95.7\%$ between fastest and slowest circuits. The remaining Opt 2 and Opt 3 results show the same general pattern across both backends and characteristics. These results show that deeper and slower circuits are much more likely to benefit from additional optimization, whereas shallow or faster circuits benefit less. 

Optimization level 3 shows a different pattern. In group 4, it helps only the Washington circuits with the longest Opt 0 runtimes, where $76.1\%$ achieve a lower total runtime. The remaining cases perform poorly, with fewer than half of the circuits in group 4 benefiting from Opt 3. On Sherbrooke, there is almost no benefit, reaching only $5\%$ for the slowest circuits and $0\%$ for the deepest. This suggests that the additional transpilation overhead of Opt 3 generally outweighs its execution time savings. For the future runtime-aware transpilation decisions, rather than using a fixed setting, optimization levels should be selected dynamically based on circuit and backend characteristics. Aggressive optimization is most beneficial for deep and slow circuits where execution-time savings are more likely to justify additional transpilation time.

\begin{table}[t]
\centering

\setlength{\tabcolsep}{2pt}

\begin{tabular}{llcccc}
\toprule
& &
\multicolumn{2}{c}{\textbf{Washington}} &
\multicolumn{2}{c}{\textbf{Sherbrooke}} \\
\cmidrule(lr){3-4}
\cmidrule(lr){5-6}

\textbf{Metric} & \textbf{Opt.}
& \textbf{Group 1} & \textbf{Group 4}
& \textbf{Group 1} & \textbf{Group 4} \\
& & (\%) & (\%) & (\%) & (\%) \\
\midrule

\multirow{3}{*}{Opt 0 exec. time}
& 1 & 15.2 & 95.7 & 20.6 & 98.6 \\
& 2 & 15.2 & 82.6 & 5.0  & 99.3 \\
& 3 & 6.5  & 76.1 & 0.0  & 5.0  \\

\midrule

\multirow{3}{*}{Circuit depth}
& 1 & 19.6 & 73.9 & 21.3 & 100.0 \\
& 2 & 13.0 & 60.9 & 9.9  & 93.6  \\
& 3 & 8.7  & 39.1 & 5.7  & 0.0   \\

\bottomrule
\end{tabular}

\caption{The impact of optimization levels on total runtime across execution-time and circuit-depth groups, measured as the percentage of circuits outperforming Opt 0.}
\label{tab:tradeoff}
\end{table}

\subsubsection{Accuracy and Inference Cost.}
The transpiled GNN achieves the strongest predictive performance among the GNN representations in most evaluated settings. Table~\ref{tab:inference} summarizes the trade-off between predictive accuracy and graph cost. The reported $R^2$ and inference time are averaged across optimization levels. Inference time was measured on a single NVIDIA RTX PRO 6000 Blackwell Server Edition GPU (batch size 8, CUDA 12.8) and three complete test set passes. The graph-size ratio represents the average node-count relative to the source graph across both backends. Graph-construction time is reported as a one-time preprocessing cost and averaged across both backends and optimization levels.

\begin{table}[t]
\centering
{\small
\setlength{\tabcolsep}{1pt}

\begin{tabular}{lcccc}
\toprule
Representation &
Mean $R^2$ &
\shortstack{Inference\\(ms/sample)} &
\shortstack{Graph-size\\ratio} &
\shortstack{Construction\\mean/median} \\
\midrule

Source GNN     & 0.619 & 0.318 & 1.0$\times$  & 149 / 47 ms \\
Hybrid GNN     & 0.629 & 0.316 & 1.0$\times$  & 149 / 47 ms \\
Transpiled GNN & \textbf{0.759} & 2.35 & 34.9$\times$ & 8040 / 149 ms \\

\bottomrule
\end{tabular}
}
\caption{Accuracy and computational cost of the GNN representations. Graph size is reported relative to the source-circuit graph.}
\label{tab:inference}
\end{table}

The Transpiled GNN processes graphs that are approximately $34.9\times$ larger on average and requires about $7.4\times$ more inference time than the other representations. However, its inference time is still in milliseconds, 2.35 ms per circuit, while achieving the highest mean $R^2$. Graph construction also introduces a large one-time preprocessing cost, although its median remains 149 ms per circuit. Overall, additional representation cost is a reasonable tradeoff for the gain in prediction accuracy.

\section{Conclusion}
This paper proposed two GNN circuit representations (hybrid and transpiled) and evaluated their prediction accuracy and overhead with a circuit dataset collected on two backends with four transpilation optimization levels.  
Our results demonstrate that the transpiled GNN generally provides the strongest performance among the graph-based representations, suggesting the importance of post-transpilation circuit structure and features. However, conventional ML regression models using transpilation-aware circuit features remain competitive with or outperform GNNs in several settings. These findings suggest that post-transpilation information is important for reliable quantum execution-time predictors, but explicit graph modeling is not always necessary. Such predictors are essential for future quantum operating systems and schedulers, where runtime estimates can guide job placement and resource allocation. 
In future work, we will evaluate a broader range of backends, noise models, simulator configurations, and transpilation strategies. 

\section*{Acknowledgments}
This work is supported in part by NSF Grant \#2426940 and a high-performance computing system acquired through NSF MRI grant DMS-1337943 to WPI.
{\small
\bibliography{aaai2027}
}
\end{document}